\documentclass{article}

\usepackage{PRIMEarxiv}

\usepackage[utf8]{inputenc} 
\usepackage[T1]{fontenc}    
\usepackage{hyperref}       
\usepackage{url}            
\usepackage{booktabs}       
\usepackage{amsfonts}       
\usepackage{nicefrac}       
\usepackage{microtype}      
\usepackage{lipsum}
\usepackage{fancyhdr}       
\usepackage{graphicx}       
\graphicspath{{media/}}     

\usepackage{cite}
\usepackage{amsmath,amssymb,amsfonts,bbm,mathtools,arydshln}
\usepackage{amsthm}
\usepackage{algorithm}
\usepackage{algpseudocode}
\usepackage{textcomp}
\usepackage{xcolor}
\usepackage{multicol}
\title{Sequence-to-Sequence Forecasting-aided State Estimation for Power Systems
\thanks{\textit{\underline{Citation}}: 
\textbf{Basulaiman, K (Basulaiman, Kamal) ; Barati, M (Barati, Masoud). Sequence-to-Sequence Forecasting-aided State Estimation for Power Systems. 2021 IEEE TEXAS POWER AND ENERGY CONFERENCE (TPEC). pages 419-424.2021. DOI:10.1109/TPEC51183.2021.9384984.}} 
}

\author{
  Kamal Basulaiman, Member, ~IEEE\\
  \textit{Industrial Engineering Department} \\
  University of Pittsburgh \\
  Pittsburgh, United States\\
  \texttt{kab391@pitt.edu} \\
   \And
  Masoud Barati, Senior Member, ~IEEE \\
  Electrical and Computer Engineering dept. \\
  University of Pittsburgh \\
  Pittsburgh, United States\\
  \texttt{masoud.barati@pitt.edu} \\
}

\begin{document}
\maketitle

\begin{abstract}
 Power system state forecasting has gained more attention in real-time operations recently. Unique challenges to energy systems are emerging with the massive deployment of renewable energy resources. As a result, power system state forecasting are becoming more crucial for monitoring, operating and securing modern power systems. This paper proposes an end-to-end deep learning framework to accurately predict multi-step power system state estimations in real-time. In our model, we employ a sequence-to-sequence framework to allow for multi-step forecasting. Bidirectional gated recurrent units (BiGRUs) are incorporated into the model to achieve high prediction accuracy. The dominant performance of our model is validated using real dataset. Experimental results show the superiority of our model in predictive power compared to existing alternatives.

\end{abstract}


\section{Introduction}
The power grid is a complex system comprising several electrical modules that transport energy from generation sites to distribution networks through transmission lines. The main objective of the power system operation and control is to consistently minimize the mismatch between power supply (generation) and demand (consumption). Smart power grids employ a set of operational devices to measure the power system variables, such as voltage magnitudes, power flows, and power injections, in real-time. Since these measurements, provided by the supervisory control and data acquisition (SCADA) system, are presumably noisy, they are transmitted to a control center in which power system state estimation (PSSE) is carried out to retrieve the correct values of the power system states. Namely, voltage magnitudes and angles at all buses of the power network. Accuracy in estimating the states of the system is critical for many operational decision making procedures in problems in power systems such as unit commitment, optimal power flow (OPF), network reconfiguration, and reliability assessment\cite{Abur:2004:1234}.

Power system state forecasting introduces unique challenges and has not received much attention in the literature. With the increasing emergence of renewable energy resources, the power system states become less predictable \cite{4b861500121845fa97e7ce5788c296ee}. Therefore, there is an urgent need for forecasting models that can accurately predict the system states ahead of time to monitor the electric power system in an economic and secure fashion.

When it comes to Power system state forecasting, there are two main streams of research in the literature. One stream revolves around the forecasting-aided state estimation (FASE) which characterizes the system state dynamics through state-transition by Kalman filtering in an online setting \cite{debs1970dynamic}. For improved extended versions, see \cite{hassanzadeh2012power} \cite{da1983state}. A recent extension was proposed by \cite{hassanzadeh2015short}, for predicting state transitions using first-order vector auto regressive (VAR) process. All approaches within this category have a strong assumption on the linearity of the state dynamics. However, such assumption breaks in reality as the dependence of system states (power flow) over time is often nonlinear.

A more recent stream of research has evolved relaxing the linear dynamics assumption by modeling state-transition as a single hidden layer in a feed-forward neural networks (FNNs) \cite{da1993state}. Making use of the universal approximation theorem \cite{csaji2001approximation}, this framework is able to capture the non-linearity in the system state dynamics \cite{do2009forecastingI} \cite{do2009forecastingII}. A major drawback though, is that the number of FNN trained parameters grows linearly with respect to the input sequence length.

To address the aforementioned limitations of FNNs, Deep RNNs were suggested by \cite{DBLP:journals/corr/abs-1811-06146} for single-step forecasting as they can capture complex long-term nonlinear dependencies in the time series data. However, RNNs in general, have their own limitations. They suffer from slow convergence in the process of capturing long-term dependencies due to possessing higher number of trainable parameters. Moreover, they suffer memory inefficiency as they store all hidden states. Lastly, for sequential multi-step forecasting, they have the tendency towards overfitting which requires creative adoption of dropout regularizers.

In this paper, we investigate the problem of obtaining accurate forecasts of multi-step-ahead power system state estimations. We advocate Gated Recurrent Units (GRUs) and show that they outperform RNNs in the context of power system states forecasting. In this work, motivated by recent advances in deep learning for sequential modeling, we propose an end-to-end sequence-to-sequence Bidirectional Gated Recurrent Unit (BiGRU) network framework for multi-step forecasting of power system state estimations. Moreover, we demonstrate the superiority of the BiGRU model against existing alternatives in the literature.


The rest of this paper is organized as follows. In section II, the underlying mathematical model is described. Section III, provides background of the recurrent models and presents the sequence-to-sequence neural network model used for multi-step power system state forecasting. In section IV, results of our computational experiments are reported. Lastly, we summarize our findings and propose directions for future research.

\textbf{Notation:}
Bold calligraphic uppercase letters denote sets or tensors e.g. $\mathcal{H}$; bold capital letters denote matrices e.g. $\mathbf{A}$, $\mathbf{B}$; bold small letters $\mathbf{a}$, $\mathbf{b}$ denote column vectors;
otherwise are scalars; 
$\dagger$ denotes the pseudo inverse; and $\times_{i}$ denotes the tensor mode product; $\odot$ denotes the element-wise  or Hadamard product; $.$ denotes the inner product.

\section{Problem Formulation}
 The problem we address is for the most part the same as that of \cite{DBLP:journals/corr/abs-1811-06146}. A brief description is provided here for the reader's convenience. Consider a graph $\mathcal{G}:= \{ \mathcal{V}, \mathcal{A}\}$, where the vertices set, $\mathcal{V}:= \{1,2,\dots, K\}$ denote the set of all buses in the grid network, and the set of arcs, $\mathcal{A}:= \{(k,j)\} \in \mathcal{V} \times \mathcal{V}$ denote the set of all transmission lines. Let $V_{k,t}$ denote the complex voltage of the $k^{th}$ bus at time $t$. $P_{k} (Q_{k})$ denotes the active (reactive) power injection. Let $P^{b}_{kj} (Q^{b}_{kj})$ denote the active (reactive) power flow observed at beginning of line $(k,j)$, and $P^{e}_{kj} (Q^{e}_{kj})$ at the end of the line $(k,j)$.
 At time $t$, the system state measurements $\mathbf{z}_{t} = $
\begin{align}
    \big[\{|V_{k,t}|^{2} \}_{k \in \mathcal{V}_{t}}, \{P_{k,t} \}_{k \in \mathcal{V}_{t}}, \{Q_{k,t} \}_{k \in \mathcal{V}_{t}}, \{P^{b}_{kj,t} \}_{(k,j) \in \mathcal{A}_{t}},\nonumber\\
    \{Q^{b}_{kj,t} \}_{(k,j) \in \mathcal{A}_{t}}, \{P^{e}_{kj,t} \}_{(k,j) \in \mathcal{A}_{t}}, \{Q^{e}_{kj,t} \}_{(k,j) \in \mathcal{A}_{t}} \big]^{T}\nonumber
\end{align} 
 are observed with the aim of recovering the system state vector $\mathbf{x}_{t} = \setlength{\dashlinegap}{2pt}
\left[\begin{array}{c:c}
\mathbf{x}^{r}_{t} & \mathbf{x}^{i}_{t}
\end{array}
\right]^{T} \, \in \, \mathbb{R}^{2K}$ from $M$ noisy measurements that, for any $t$, adhere to the equation
 \begin{equation}
 \mathbf{z}_{t} = h_{t}(\mathbf{x}_{t}) + \boldsymbol{\varepsilon}_{t} = \mathcal{H}_{t} \, \times_1 \, \mathbf{x}_{t} \, \times_2 \, \mathbf{x}_{t} + \boldsymbol{\varepsilon}_{t} \label{eq1}
 \end{equation}
  where for a fixed $t$, $\mathbf{z}_{t}, \, \boldsymbol{\varepsilon}_{t} \, \in \, \mathbb{R}^{M}$ and the measurement tensor $\mathcal{H}_{t}\, \in \, \mathbb{R}^{2K \times 2K \times M}$. For more details on tensor mode product, see \cite{almutairi2019prema}.

\begin{figure}[t]
\centerline{\includegraphics[scale=0.28]{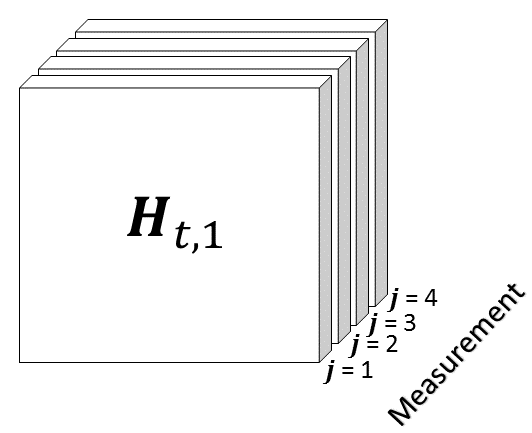}}
\caption{For $\mathcal{H}_{t}$ with $M = 4$, each slice is a matrix $\mathbf{H}_{t, j}\, \in \, \mathbb{R}^{2K \times 2K}$.}
\label{fig1}
\end{figure}
 
\section{Preliminaries}
Recurrent models are designed to capture nonlinear dependencies exhibited in time series data. In general, a recurrent model can be seen as a nonlinear dynamical system with a differentiable state transition function $f_{\theta}: \mathbb{R}^{n} \times \mathbb{R}^{d} \rightarrow \mathbb{R}^{n}$, where $\mathbf{x}_{t} \in \mathbb{R}^{d}$ is the input to the system at time $t$, $\mathbf{\theta} \in \mathbb{R}^{m}$ is the parameter vector, and $\mathbf{h}_{t} \in \mathbb{R}^{n}$ is the internal or hidden state of the discrete-time system which follow \cite{Chen16p} \cite{DBLP:journals/corr/abs-1808-03314} \cite{DBLP:journals/corr/abs-1805-10369}:
\begin{align*}
    \mathbf{h}_{t} = f_{\theta}(\mathbf{h}_{t-1}, \mathbf{x}_{t})
\end{align*}
\subsection{Recurrent Neural Network (RNN)}
Given some weight matrices $\mathbf{W} \in \mathbb{R}^{n \times n}$, $\mathbf{U} \in \mathbb{R}^{n \times d}$, and some nonlinear function $\varphi$, the state-transition function for a recurrent neural network is \cite{Chen16p} \cite{DBLP:journals/corr/abs-1808-03314} \cite{DBLP:journals/corr/abs-1805-10369}:
\begin{align*}
    \mathbf{h}_{t} = \varphi(\mathbf{W} \, \mathbf{h}_{t-1} + \mathbf{U} \, \mathbf{x}_{t})
\end{align*}

\subsection{Gated Recurrent Unit (GRU)}
GRU networks are well-known class of recurrent models. Let the internal state vector be the vector $\mathbf{s} = \mathbf{h}$. Then an GRU layer has the following weights :
\begin{itemize}
    \item Recurrent weights: $\mathbf{W}_{zx}, \mathbf{W}_{rx}, \mathbf{W}_{x}, \in \, \mathbb{R}^{d \times d}$
    \item Input weights: $\mathbf{W}_{zh}, \mathbf{W}_{rh}, \mathbf{W}_{h}, \mathbf{W}_{oh} \, \in \, \mathbb{R}^{d \times n}$
    \item Bias weights: $\mathbf{b}_{z}, \mathbf{b}_{r}, \mathbf{b}, \mathbf{b}_{o} \, \in \, \mathbb{R}^{n}$
\end{itemize}
 The state-transition function of the GRU (a layer forward pass) can be written as \cite{DBLP:journals/corr/ChungGCB14}:
\begin{align*}
    \mathbf{z}_{t} &= \sigma(\mathbf{W}_{zx} \, . \, \mathbf{x}_{t} + \mathbf{W}_{zh} \, . \, \mathbf{h}_{t-1} + \mathbf{b}_{z})\\
    \mathbf{r}_{t} &= \sigma(\mathbf{W}_{rx} \, . \, \mathbf{x}_{t} + \mathbf{W}_{rh} \, . \, \mathbf{h}_{t-1} + \mathbf{b}_{r})\\
    \mathbf{\Tilde{h}}_{t} &= tanh(\mathbf{W}_{h} \, ( \mathbf{r}_{t} \odot \mathbf{h}_{t-1}) + \mathbf{W}_{x} \, . \, \mathbf{x}_{t} + \mathbf{b})\\
    \mathbf{h}_{t} &= (1-\mathbf{z}_{t}) \, \odot \, \mathbf{h}_{t-1}  + \mathbf{z}_{t} \odot \mathbf{\Tilde{h}}_{t}\\
    \mathbf{o}_{t} &= \mathbf{W}_{oh} \, . \, \mathbf{h}_{t} + \mathbf{b}_{o}\\
    \hat{\mathbf{y}}_{t} &= \mathbf{o}_{t}
\end{align*}
Where $\mathbf{z}_{t}$, $\mathbf{r}_{t}$, $\mathbf{\Tilde{h}}_{t}$, $\mathbf{h}_{t}$, $\mathbf{o}_{t}$ are update activation, rese gate, candidate activation, hidden (internal) state and output gate respectively. $\sigma$ is a point-wise non-linear activation function (e.g. sigmoid function). $\hat{\mathbf{y}}_{t}$ is the predicted output at time $t$.
\begin{figure}[t]
\includegraphics[width=\columnwidth]{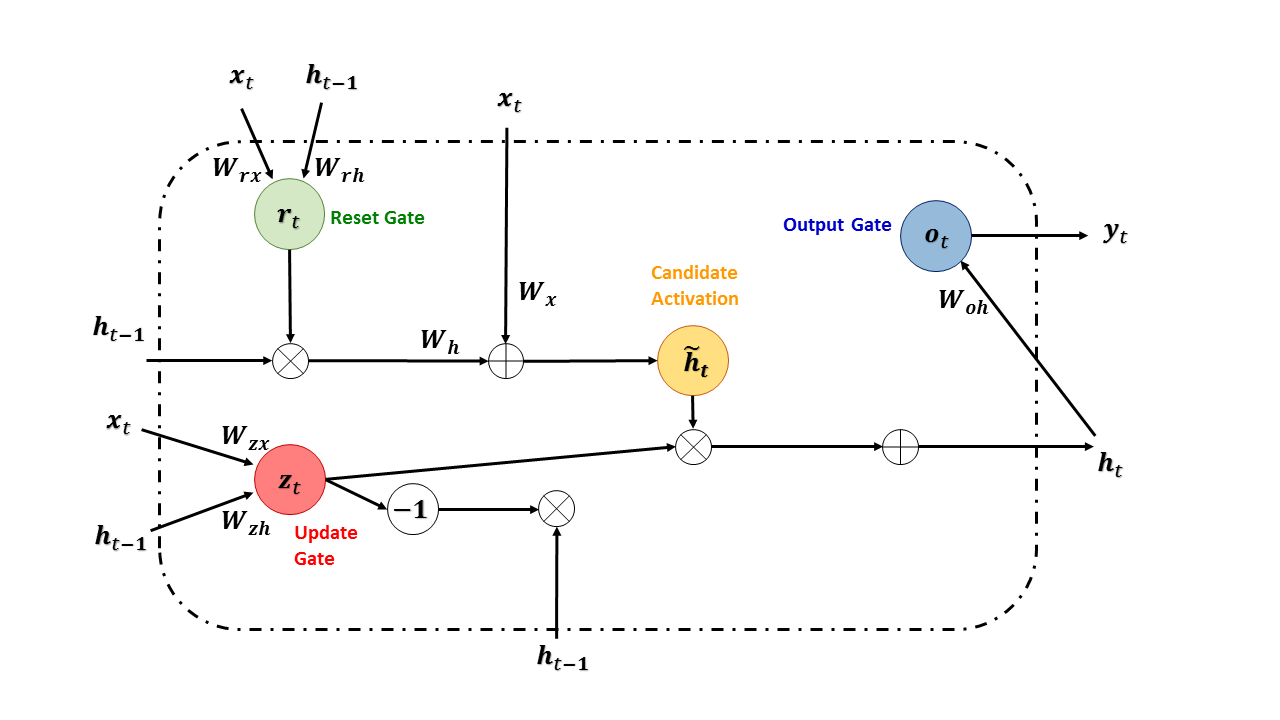}
\caption{Detailed structure of the Gated Recurrent Unit cell.}
\label{fig2}
\end{figure}
After unfolding the GRU unit, shown in fig \ref{fig3}, computing the error back-propagation with least square objective function is carried out as follows:
\begin{align*}
    \mathcal{L}(\mathbf{x}, \theta) = \sum\limits_{t} \, \frac{1}{2} (y_{t} - \hat{y}_{t})^{2}
\end{align*}
where $\boldsymbol{\theta}$ contains all recurrent and input weight matrices with folded biases into them. At time $t$, the back-propagation  is described as: 
\begin{align*}
    d\hat{\mathbf{y}}_{t} &= \frac{\partial \mathcal{L}(t)}{\partial \hat{\mathbf{y}}_t} = \mathbf{y}_{t} - \hat{\mathbf{y}}_{t} \\
    d\mathbf{h}_{t} &= \frac{\partial \mathcal{L}(t)}{\partial \mathbf{h}_t} = \frac{\partial \mathcal{L}(t)}{\partial \hat{\mathbf{y}}_t} = \frac{\partial \hat{\mathbf{y}}_t}{\partial \mathbf{h}_t} = (\mathbf{y}_{t} - \hat{\mathbf{y}}_{t})^{T} \mathbf{W}_{oh} = \delta \\
    d\Tilde{\mathbf{h}}_{t} &= \delta \odot \mathbf{z}_t\\
    d\mathbf{W}_{x} &= \left(\delta \odot \mathbf{z}_{t} \odot dtanh \right)^{T} \mathbf{x}_{t}\\
    d\mathbf{W}_{h} &= \left(\delta \odot \mathbf{z}_{t} \odot dtanh \right)^{T} (\mathbf{r}_{t} \odot \mathbf{h}_{t-1})\\
    d\mathbf{W}_{rx} &= \Big[\left(\delta \odot \mathbf{z}_{t} \odot dtanh \right)^{T} \mathbf{W}_{h} \odot d\sigma_{r} \odot \mathbf{h}_{t-1}\Big]^{T} \mathbf{x}_t\\
    d\mathbf{W}_{rh} &= \Big[\left(\delta \odot \mathbf{z}_{t} \odot dtanh \right)^{T} \mathbf{W}_{h} \odot d\sigma_{r} \odot \mathbf{h}_{t-1}\Big]^{T} \mathbf{h}_{t-1}\\
    d\mathbf{W}_{zx} &= \left(\delta \odot (\Tilde{\mathbf{h}}_{t} - \mathbf{h}_{t}) \odot d\sigma_{z} \right)^{T} \mathbf{x}_{t}\\
    d\mathbf{W}_{zh} &= \left(\delta \odot (\Tilde{\mathbf{h}}_{t} - \mathbf{h}_{t}) \odot d\sigma_{z} \right)^{T} \mathbf{h}_{t-1}\\
    d\mathbf{x}_{t} &= \Big[\left(\delta \odot \mathbf{z}_{t} \odot dtanh \right)^{T} \mathbf{W}_{h} \odot \mathbf{h}_{t-1} \odot d\sigma_{r} \Big]^{T} \mathbf{W}_{rx} \\
    & + \left(\delta \odot \mathbf{z}_{t} \odot dtanh \right)^{T} \mathbf{W}_{x} \\
    & + \left(\delta \odot (\Tilde{\mathbf{h}}_{t} - \mathbf{h}_{t}) \odot d\sigma_{z} \right)^{T} \mathbf{W}_{zx}\\
    d\mathbf{h}_{t-1} &= \Big[\left(\delta \odot \mathbf{z}_{t} \odot dtanh \right)^{T} \mathbf{W}_{h} \odot \mathbf{h}_{t-1} \odot d\sigma_{r} \Big]^{T} \mathbf{W}_{rh} \\
    & + \left(\delta \odot \mathbf{z}_{t} \odot dtanh \right)^{T} (\mathbf{W}_{h} \odot \mathbf{r}_t)\\
    & + \left(\delta \odot (\Tilde{\mathbf{h}}_{t} - \mathbf{h}_{t}) \odot d\sigma_{z} \right)^{T} \mathbf{W}_{zh} + \delta \odot (1 - \mathbf{z}_t)\\
\end{align*}
Where, $d\tanh$, $d\sigma_{r}$, and $d\sigma_{z}$ represent the derivatives of the piece-wise nonlinear activation functions $tanh(.)$ and $\sigma(.)$ contained in $\Tilde{\mathbf{h}}_t$, $\mathbf{r}_t$, and $\mathbf{z}_t$ respectively. 

\subsection{Bidirectional GRU (BiGRU)}
A variation of the typical GRU explained above is the bidirectional GRU which takes into consideration the relationship between the previous output the current and the subsequent outputs. \cite{zhao2018deep}. A good example is used in the field of natural language processing (NLP). More precisely, in text translation the context of the sentence is crucial for predicting the next word, which in turn, can be strongly related to previous words as well as words that come later in the sentence.
In the Bidirectional setting, the forward GRU passes through the input sequence in order from $\mathbf{x}_1$ to $\mathbf{x}_T$ and produces a hidden forward sequence $\overrightarrow{\mathbf{h}}_t$, while the backward GRU passes through the input in reverse order and computes a backward hidden sequence $\overleftarrow{\mathbf{h}}_t$. The two sequences are then considered together to compute the output \cite{zhao2018deep}:
\begin{align*}
    \overrightarrow{\mathbf{h}}_t&= (1-\mathbf{z}_{t}) \, \odot \, \overrightarrow{\mathbf{h}}_{t-1}  + \mathbf{z}_{t} \odot \overrightarrow{\mathbf{\Tilde{h}}}_{t}\nonumber \\
    \overleftarrow{\mathbf{h}}_t&= (1-\mathbf{z}_{t}) \, \odot \, \overleftarrow{\mathbf{h}}_{t-1}  + \mathbf{z}_{t} \odot \overleftarrow{\mathbf{\Tilde{h}}}_{t}\nonumber \\
    \mathbf{o}_{t} &= \mathbf{W}_{o\overrightarrow{h}} \, . \, \overrightarrow{\mathbf{h}}_{t} + \mathbf{W}_{o\overleftarrow{h}} \, . \, \overleftarrow{\mathbf{h}}_{t} + \mathbf{b}_{o} \nonumber\\
    \mathbf{\hat{y}}_t&=\mathbf{o}_t
\end{align*}
\subsection{GRU with 1D Convolutional Layer (1DConvGRU)}
\begin{figure}[t]
\centerline{\includegraphics[width=0.9\columnwidth]{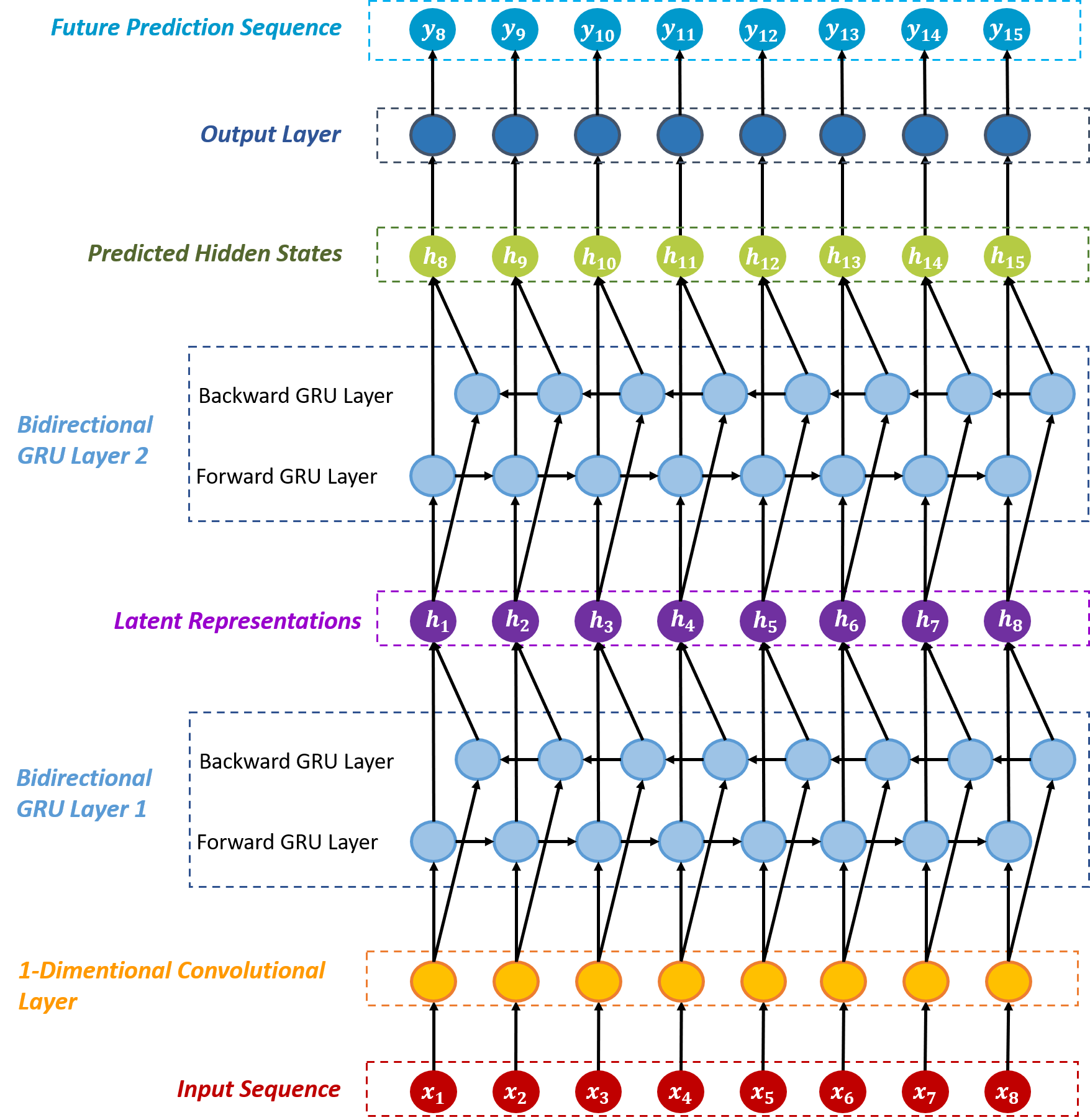}}
\caption{Illustration of the architecture of our BiGRU model with sequence length of size 8.}
\label{fig3}
\end{figure}

Here we adopt the idea of representation learning where feature engineering of the input data is learned along with the other parameters of the GRU networks. Convolutional Neural Networks (CNN) have accomplished tremendous success in language processing, object detection and image and recognition \cite{GoodBengCour16}. It was due the unique structure it possess, namely, convolutions, the ability to divide the original task into smaller ones and easily analyze them become possible. In Conv1D, convolutions are filters with one dimensional (time) windows that slide over the time series \cite{DBLP:journals/corr/abs-1809-04356}. Those filters can also be seen as some non-linear transformations of a time series inputs \cite{DBLP:journals/corr/abs-1809-04356} \cite{GoodBengCour16}. Generally, the convolution at time $t$ is
\begin{align*}
    \mathbf{C}_{t} = \boldsymbol{\varphi}(\boldsymbol{\omega} * \mathbf{x}_{t-\frac{l}{2} : t+\frac{l}{2}} + \mathbf{b}) \hspace{7 mm} \forall t \in [1,T]
\end{align*}
where $\mathbf{C}$ denotes the convoluted result out of applying  the dot product to the univariate time series $\mathbf{x}$ of length $T$ with a filter $\boldsymbol{\omega}$ of length $l$, a bias parameter $\mathbf{b}$ and a non-linear activation function $\boldsymbol{\varphi}$. The output of one convolution $\mathbf{C}$ applied on an input time series $\mathbf{x}$ is indeed nothing but a filtered univariate time series \cite{GoodBengCour16}. So, multiple filters when applied on a time series will produce a multivariate time series with dimensions equal to the number of filters used \cite{DBLP:journals/corr/abs-1809-04356}. Applying several filters on a time series input would be equivalent to learn significant discriminative features for forecasting such as seasonality and trends.
Interestingly enough, $\forall \ t \in [1, T]$ the result can be found using the same convolution (the same filter values $\boldsymbol{\omega}$ and $\mathbf{b}$). This property of the CNN is known as weight sharing. It enables the CNN to learn time-invariant filters \cite{DBLP:journals/corr/abs-1809-04356}. We combined the 1D convolutional layer, which will do the feature engineering, with bidirectional GRU layers, see figure \ref{fig3} for futher details, which will do the forecasting, to make the best of the two models.
\section{Experimental Results}
In this section, computational experiments were conducted to evaluate and compare the performance of our (BiGRU) model to the state-of-the-art in power system state forecasting used by \cite{DBLP:journals/corr/abs-1811-06146}. We first describe the dataset used in this study as well as the experimental setup, and we  then discuss the results in details.

\subsection{Experiment Design}
The focus of this experiment is to empirically and fairly compare different deep recurrent NN models, our model against the state-of-the-art model proposed in \cite{DBLP:journals/corr/abs-1811-06146}. Hence, our experiments are designed to keep the setup simple and the comparisons fair. All numerical computations were carried out on google Colabratory GPU.
\texttt{Keras} and \texttt{Tensorflow} as a backend \cite{abadi2016tensorflow} were used to train all neural networks using the Adaptive moment estimation optimizer \texttt{Adam} (back propagation variant) with initial learning rate of $10^{-3}$ for $100$ epochs. Also, the depth of all neural network models was fixed to $3$ hidden layers. In addition, the activation functions were fixed to ReLU. For the 1DConv layer, number of filters was set to $64$, kernal size of $5$, stride of $1$, and no paddings. This essentially translates into extracting $64$ features out of the original input sequence. The length of the input sequence were initially set equal to $5$ and then were assigned different values for sensitivity analysis purposes. Furthermore, dropouts of $5\%$ were added to all neural networks in order to avoid overfitting. Lastly, all experiments were repeated for $30$ times and the outputs were averaged out and reported for statistical consistency purposes.

\begin{figure}[t]
\centerline{\includegraphics[width=0.9\columnwidth]{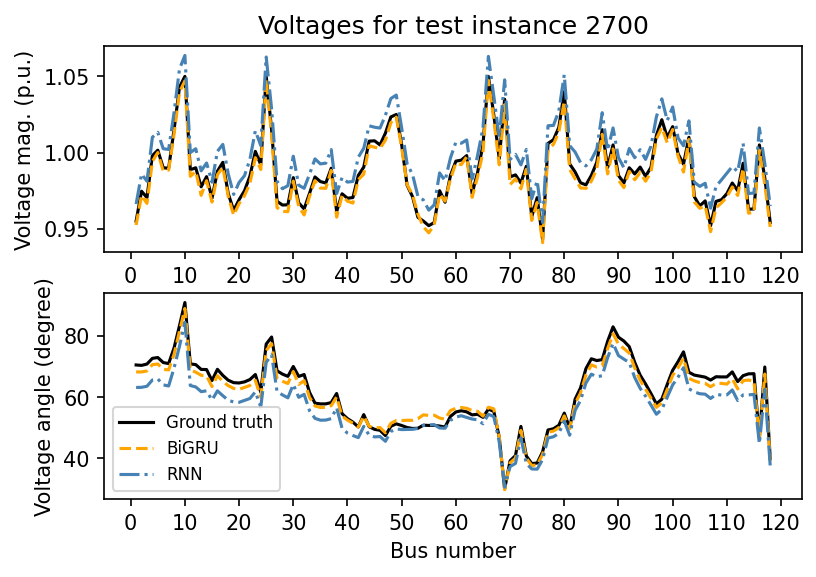}}
\centerline{\includegraphics[width=0.9\columnwidth]{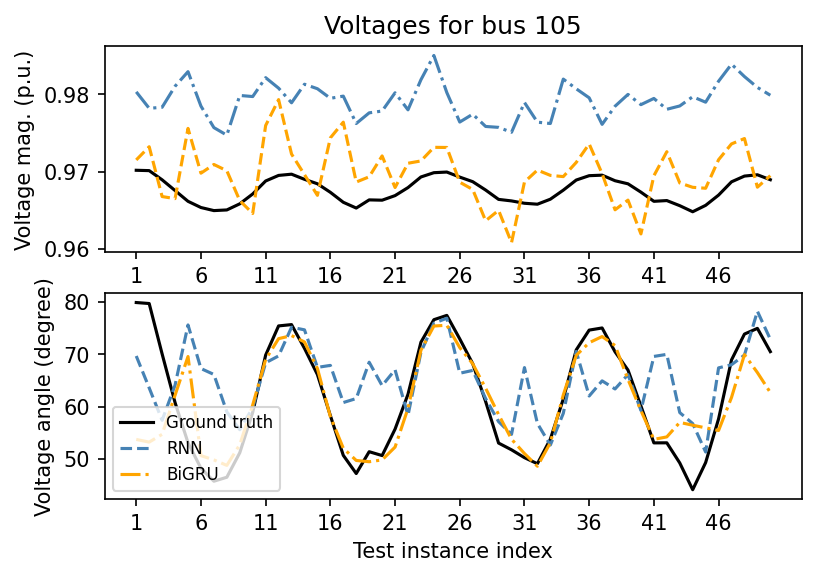}}
\caption{Sequence length = 5. (above) prediction error in voltage magnitude and angels of 118-bus system at future time 2700, and (below) prediction error at bus 50 of the 118-bus system for future time 1 to 50.}
\label{fig4}
\end{figure}

\subsection{Dataset}
We used the GEFCom2014 Load Forecasting dataset provided in \cite{HONG2016896} and MATPOWER to generate a IEEE 118-bus system dataset. It has $20,000$ time instances of complex voltage readings $\{\mathbf{z}_{t}, \mathbf{x}_{t}\}_{t=1}^{20,000}$. We split it into three parts: $75\%$ as training set, $5\%$ as validation set used with 5-fold cross-validation for early stopping and for optimizing the hyperparameters, and $20\%$ as test set for the final evaluation.

\subsection{Results}
This section examines the predictive performance of our BiGRU model and compare it to the RNN proposed by \cite{DBLP:journals/corr/abs-1811-06146}. Both models forecast a sequence of future power system states in milliseconds.
For each model, an experiment is conducted using the ground-truth for training under four different values of the sequence length $l =\{5, 10, 15, 20\}$. The prediction accuracy of each model was measured and evaluated using the normalized root mean square error (NRMSE) performance measure on the test set.
\begin{figure}[t]
\centerline{\includegraphics[width=0.9\columnwidth]{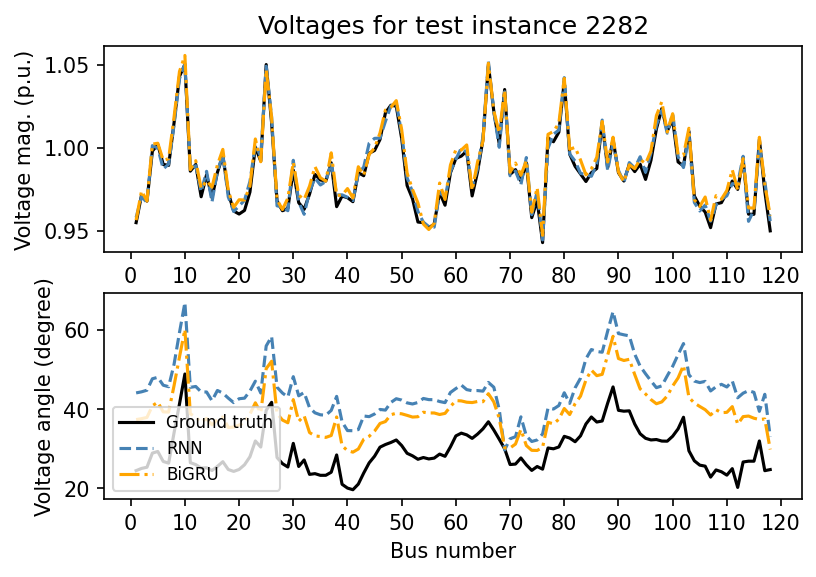}}
\centerline{\includegraphics[width=0.9\columnwidth]{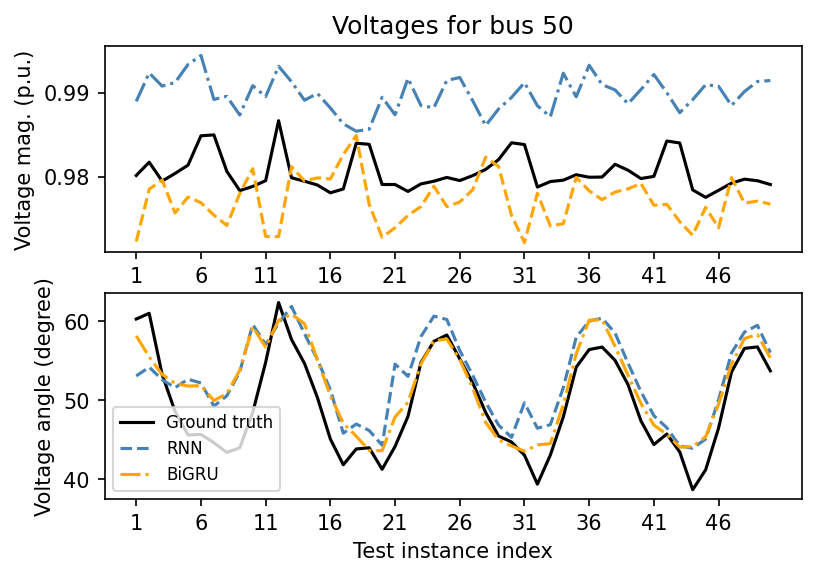}}
\caption{Sequence length = 10. (above) prediction error in voltage magnitude and angels of 118-bus system at future time 2700, and (below) prediction error at bus 50 of the 118-bus system for future time 1 to 50.}
\label{fig5}
\end{figure}

\begin{figure}[t]
\centerline{\includegraphics[width=0.9\columnwidth]{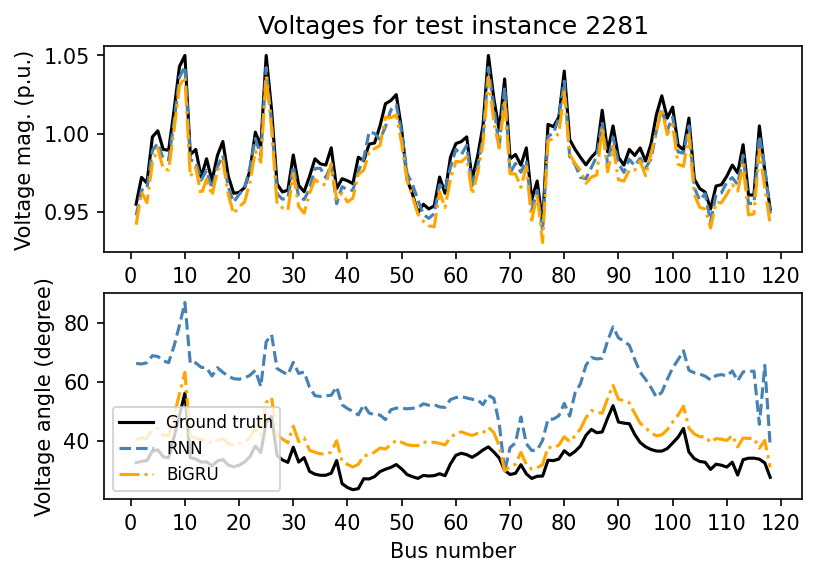}}
\centerline{\includegraphics[width=0.9\columnwidth]{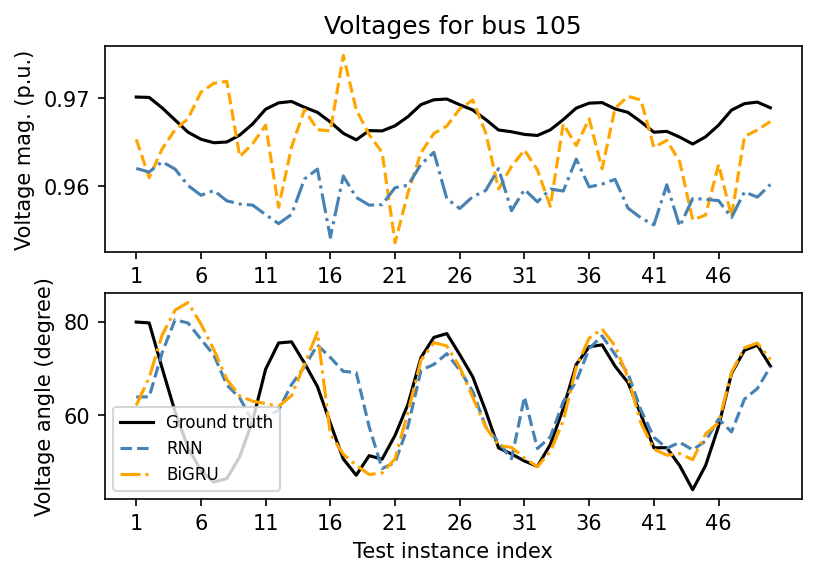}}
\caption{Sequence length = 15. (above) prediction error in voltage magnitude and angels of 118-bus system at future time 2700, and (below) prediction error at bus 50 of the 118-bus system for future time 1 to 50.}
\label{fig6}
\end{figure}
We  begin  with  comparisons in Table I which shows the predictive performance, based on the NRMSE metric,  of our model and the baseline used by \cite{DBLP:journals/corr/abs-1811-06146}, evaluated over the entire test dataset. 
\begin{table}[t]
\caption{Forecasting performance comparison of all NN models with different input length}
\begin{center}
\begin{tabular}{|c|c|c|c|c|}
\hline
\textbf{NN}&\multicolumn{4}{|c|}{\textbf{NRMSE}} \\
\cline{2-5} 
\textbf{Model} & \textbf{\textit{$l=5$}}& \textbf{\textit{$l=10$}}& \textbf{\textit{$l=15$}} & \textbf{\textit{$l=20$}} \\
\hline
BiGRU & $\mathbf{0.005485}$ & $\mathbf{0.005493}$ & $\mathbf{0.005414}$ & $\mathbf{0.006017}$\\
RNN & 0.008208 & 0.006125 & 0.007477 & 0.008415\\
\hline
\end{tabular}
\label{tab1}
\end{center}
\end{table}
The performance of RNN decays as the sequence length increases which validates the limitation of slow convergence in RNNs. This is due the phenomenon of vanishing/exploding gradient and the fact that RNNs tend to overfit when no dropout is used. On the contrary, BiGRU outperforms RNN under all sequence lengths $l$ by a significant magnitude. This due many factors: (1) the structured gates which help avoiding the problem of vanishing/exploding gradient. (2) The bidirectional structure allows for learning from both previous and subsequent data points which boosts the accuracy and learning stability. As a result, our BiGRU model was able to achieve similar performance under different sequence lengths.

In figure \ref{fig4}, the true voltages and their forecasts provided by the RNN, and BiGRU, with sequence length of $l = 5$, for all buses on future time step $2700$ are reported in the upper part and for bus 105 of the 118-bus system for future time steps 1 to 50 are reported in the lower part of the figure.
The ground-truth and forecast voltages provided by the RNN, and BiGRU, with sequence length of $l = 10$, for all buses of the 118-bus system on future time step $2282$ as well as first 50 future time steps at bus $50$ are depicted in figure \ref{fig5}.
In figure \ref{fig6}, the true voltages and their forecasts provided by the RNN, and BiGRU, with sequence length of $l = 15$, for all buses on future time step $2281$ are depicted in the upper part and for bus 105 of the 118-bus system for future time steps 1 to 50 are depicted in the lower part of the figure.
Finally, the ground-truth and forecast voltages provided by the RNN, and BiGRU, with sequence length of $l = 20$, for all buses of the 118-bus system on future time step $2281$ as well as first 50 future time steps at bus $105$ are depicted in figure \ref{fig5}.
The choice of future time steps $2700, 2281, 2282$ and buses $50, 105$ in the figures was based on the variance of these vectors. As it is well-known that vectors with the highest variance are the most challenging to fit. The plots in figures $4 - 7$ illustrate that our BiGRU model performs the best in all cases against the RNN used in \cite{DBLP:journals/corr/abs-1811-06146}.

\begin{figure}[t]
\centerline{\includegraphics[width=0.9\columnwidth]{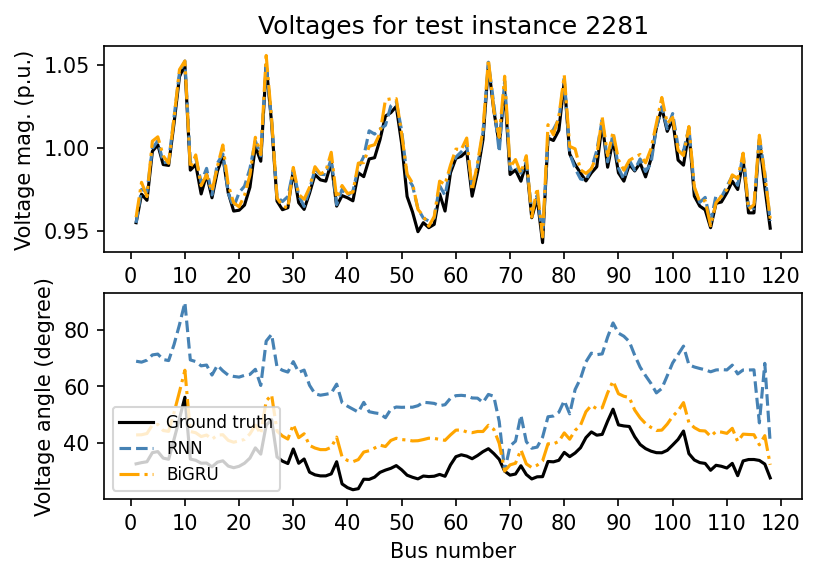}}
\centerline{\includegraphics[width=0.9\columnwidth]{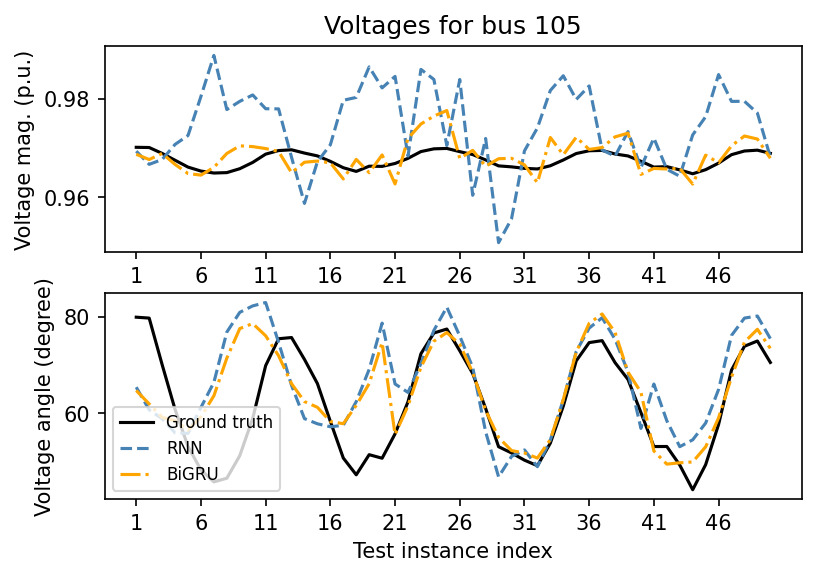}}
\caption{Sequence length = 20. (above) prediction error in voltage magnitude and angels of 118-bus system at future time 2700, and (below) prediction error at bus 50 of the 118-bus system for future time 1 to 50.}
\label{fig7}
\end{figure}
\section{Conclusion}
This paper addresses the task of accurately forecasting a sequence of the power system states. Toward that goal, we proposed an end-to-end sequence-to-sequence deep learning model. Our model leverages the strength of 1DConv layers in feature extraction, GRU in capturing long-term dependencies, and bidirectional structure for powerful representation and stability. The proposed model have high predictive power (accuracy) and strong performance stability in capturing the long-term nonlinear dependencies under different sequence lengths. Our  experiments on the IEEE 118-bus benchmark system demonstrate that our model outperforms the baseline. This work can be further extended to forecast power system states where corruptions are present in the input data.



\vspace{12pt}
\newpage
\bibliographystyle{unsrt}
\bibliography{ref.bib}

\begin{thebibliography}{10}

\bibitem{Abur:2004:1234}
A.~Abur and A.~G. Exposito.
\newblock {\em Power system state estimation: theory and implementation}.
\newblock CRC press, 2004.

\bibitem{4b861500121845fa97e7ce5788c296ee}
Yingchen Zhang, Rui Yang, Jie Zhang, Yang Weng, and {Bri Mathias} Hodge.
\newblock {\em Predictive Analytics for Comprehensive Energy Systems State
  Estimation}, pages 343--376.
\newblock Elsevier, 11 2017.

\bibitem{debs1970dynamic}
Atif~S Debs and Robert~E Larson.
\newblock A dynamic estimator for tracking the state of a power system.
\newblock {\em IEEE Transactions on Power Apparatus and Systems},
  (7):1670--1678, 1970.

\bibitem{hassanzadeh2012power}
Mohammad Hassanzadeh and Cans{\i}n~Yaman Evrenoso{\u{g}}lu.
\newblock Power system state forecasting using regression analysis.
\newblock In {\em 2012 IEEE Power and Energy Society General Meeting}, pages
  1--6. IEEE, 2012.

\bibitem{da1983state}
AM~Leite Da~Silva, MB~Do~Coutto~Filho, and JF~De~Queiroz.
\newblock State forecasting in electric power systems.
\newblock In {\em IEE Proceedings C (Generation, Transmission and
  Distribution)}, volume 130, pages 237--244. IET, 1983.

\bibitem{hassanzadeh2015short}
Mohammad Hassanzadeh, Cans{\i}n~Yaman Evrenoso{\u{g}}lu, and Lamine Mili.
\newblock A short-term nodal voltage phasor forecasting method using temporal
  and spatial correlation.
\newblock {\em IEEE Transactions on Power Systems}, 31(5):3881--3890, 2015.

\bibitem{da1993state}
AP~Alves da~Silva, AM~Leite da~Silva, JC~Stacchini de~Souza, and
  MB~Do~Coutto~Filho.
\newblock State forecasting based on artificial neural networks.
\newblock In {\em Proc. 11th PSCC}, pages 461--467, 1993.

\bibitem{csaji2001approximation}
Bal{\'a}zs~Csan{\'a}d Cs{\'a}ji.
\newblock Approximation with artificial neural networks.
\newblock {\em Faculty of Sciences, Etvs Lornd University, Hungary}, 24:48,
  2001.

\bibitem{do2009forecastingI}
Milton~Brown Do~Coutto~Filho and Julio Cesar~Stacchini de~Souza.
\newblock Forecasting-aided state estimation—part i: Panorama.
\newblock {\em IEEE Transactions on Power Systems}, 24(4):1667--1677, 2009.

\bibitem{do2009forecastingII}
Milton~Brown Do~Coutto~Filho, Julio Cesar~Stacchini de~Souza, and
  Ronaldo~S{\'E}rgio Freund.
\newblock Forecasting-aided state estimation—part ii: Implementation.
\newblock {\em IEEE Transactions on Power Systems}, 24(4):1678--1685, 2009.

\bibitem{DBLP:journals/corr/abs-1811-06146}
Liang Zhang, Gang Wang, and Georgios~B. Giannakis.
\newblock Real-time power system state estimation and forecasting via deep
  neural networks.
\newblock {\em CoRR}, abs/1811.06146, 2018.

\bibitem{almutairi2019prema}
Faisal~M. Almutairi, Charilaos~I. Kanatsoulis, and Nicholas~D. Sidiropoulos.
\newblock Prema: Principled tensor data recovery from multiple aggregated
  views, 2019.

\bibitem{Chen16p}
Gang Chen.
\newblock A gentle tutorial of recurrent neural network with error
  backpropagation.
\newblock {\em CoRR}, abs/1610.02583, 2016.

\bibitem{DBLP:journals/corr/abs-1808-03314}
Alex Sherstinsky.
\newblock Fundamentals of recurrent neural network {(RNN)} and long short-term
  memory {(LSTM)} network.
\newblock {\em CoRR}, abs/1808.03314, 2018.

\bibitem{DBLP:journals/corr/abs-1805-10369}
John Miller and Moritz Hardt.
\newblock When recurrent models don't need to be recurrent.
\newblock {\em CoRR}, abs/1805.10369, 2018.

\bibitem{DBLP:journals/corr/ChungGCB14}
Junyoung Chung, {\c{C}}aglar G{\"{u}}l{\c{c}}ehre, KyungHyun Cho, and Yoshua
  Bengio.
\newblock Empirical evaluation of gated recurrent neural networks on sequence
  modeling.
\newblock {\em CoRR}, abs/1412.3555, 2014.

\bibitem{zhao2018deep}
Yu~Zhao, Rennong Yang, Guillaume Chevalier, Ximeng Xu, and Zhenxing Zhang.
\newblock Deep residual bidir-lstm for human activity recognition using
  wearable sensors.
\newblock {\em Mathematical Problems in Engineering}, 2018, 2018.

\bibitem{GoodBengCour16}
Ian~J. Goodfellow, Yoshua Bengio, and Aaron Courville.
\newblock {\em Deep Learning}.
\newblock MIT Press, Cambridge, MA, USA, 2016.
\newblock \url{http://www.deeplearningbook.org}.

\bibitem{DBLP:journals/corr/abs-1809-04356}
Hassan~Ismail Fawaz, Germain Forestier, Jonathan Weber, Lhassane Idoumghar, and
  Pierre{-}Alain Muller.
\newblock Deep learning for time series classification: a review.
\newblock {\em CoRR}, abs/1809.04356, 2018.

\bibitem{abadi2016tensorflow}
Mart{\'\i}n Abadi, Ashish Agarwal, Paul Barham, Eugene Brevdo, Zhifeng Chen,
  Craig Citro, Greg~S Corrado, Andy Davis, Jeffrey Dean, Matthieu Devin, et~al.
\newblock Tensorflow: Large-scale machine learning on heterogeneous distributed
  systems.
\newblock {\em arXiv preprint arXiv:1603.04467}, 2016.

\bibitem{HONG2016896}
Tao Hong, Pierre Pinson, Shu Fan, Hamidreza Zareipour, Alberto Troccoli, and
  Rob~J. Hyndman.
\newblock Probabilistic energy forecasting: Global energy forecasting
  competition 2014 and beyond.
\newblock {\em International Journal of Forecasting}, 32(3):896 -- 913, 2016.

\end{thebibliography}

\end{document}